# Infrared Measurement of the Pseudogap in P-Doped and Co-Doped BaFe$_2$As$_2$ High-Temperature Superconductors

S. J. Moon,[1,2] A. A. Schafgans,[1] S. Kasahara,[3] T. Shibauchi,[4] T. Terashima,[3] Y. Matsuda,[4] M. A. Tanatar,[5] R. Prozorov,[5] A. Thaler,[5] P. C. Canfield,[5] A. S. Sefat,[6] D. Mandrus,[6,7] and D. N. Basov[1,*]

[1]Department of Physics, University of California, San Diego, La Jolla, California 92093, USA
[2]Department of Physics, Hanyang University, Seoul 133-791, South Korea
[3]Research Center for Low Temperature and Materials Science, Kyoto University, Kyoto 606-8502, Japan
[4]Department of Physics, Kyoto University, Kyoto 606-8502, Japan
[5]Ames Laboratory and Department of Physics and Astronomy, Iowa State University, Ames, Iowa 50011, USA
[6]Materials Science and Technology Division, Oak Ridge National Laboratory, Oak Ridge, Tennessee 37831, USA
[7]Department of Materials Science and Engineering, University of Tennessee, Knoxville, Tennessee 37996, USA

We report on infrared studies of charge dynamics in a prototypical pnictide system: the BaFe$_2$As$_2$ family. Our experiments have identified hallmarks of the pseudogap state in the BaFe$_2$As$_2$ system that mirror the spectroscopic manifestations of the pseudogap in the cuprates. The magnitude of the infrared pseudogap is in accord with that of the spin-density-wave gap of the parent compound. By monitoring the superconducting gap of both P- and Co-doped compounds, we find that the infrared pseudogap is unrelated to superconductivity. The appearance of the pseudogap is found to correlate with the evolution of the antiferromagnetic fluctuations associated with the spin-density-wave instability. The strong-coupling analysis of infrared data further reveals the interdependence between the magnetism and the pseudogap in the iron pnictides.

The nature of the pseudogap phase and its relation to high-transition-temperature ($T_c$) cuprate remains a significant yet unresolved problem in condensed matter physics [1]. While the pseudogap in the cuprates has been extensively documented [1, 2], there has been only circumstantial experimental evidence for the existence of a pseudogap in the iron pnictides [3-7]. Similar with the cuprates, high-$T_c$ superconductivity in the iron pnictides emerges when the antiferromagnetic (AFM) order of parent compounds is suppressed, for example, via chemical substitution. The electronic ground states of the respective parent phases, however, are qualitatively different: Mott insulator in the cuprates and bad metals in the pnictides [8-10]. In the phase diagram of the cuprates, there are at least two distinct energy gaps: one due to Mott localization of AFM parent compounds and the other due to superconductivity. Among many scenarios, the pseudogap in the cuprates has been discussed in direct connection with these two gaps as either the persistence of the parent compound gap in the doped phases or the precursor of the superconducting gap above $T_c$.

In this Letter, we investigate the electronic response of the BaFe$_2$As$_2$ (Ba122) system using infrared spectroscopy. Our data not only establish the pseudogap in this family of the iron pnictides but also reveal the infrared manifestations of the pseudogap that are quite similar with those in the cuprates [2]. By exploring the evolution of the charge dynamics across the phase diagram we are able to narrow down the field of plausible scenarios of the pseudogap state. Specifically, the superconducting gap and the pseudogap in P-doped and Co-doped Ba122 superconductors lead to two distinct spectroscopic features suggesting that the infrared pseudogap in both of these compounds is not related to superconductivity.

Figures 1(a)-1(d) show the real part of optical conductivity $\sigma_1(\omega)$ of BaFe$_2$(As$_{0.67}$P$_{0.33}$)$_2$ (P-Ba122) and Ba(Fe$_{1-x}$Co$_x$)$_2$As$_2$ (Co-Ba122) at various temperatures ($T$) (see the Supplemental Material [11]). New insights into the electromagnetic response of this series of materials are provided by the data for P-Ba122 and overdoped (OD) Co-Ba122 ($x$=0.25) that have not been investigated using infrared spectroscopy in the past. Furthermore, the P-Ba122 system sets the standard for a superconductor with low disorder, as attested by quantum oscillations experiments [12]. We supplement these new data with the results for optimally doped (OPD) Co-Ba122 ($x$=0.08) and parent compound that agree with the published works [13-20]. We find that $\sigma_1(\omega)$ of all the compounds is quite similar at 295 K. The far-infrared response is dominated by a narrow Drude-like feature followed by a smooth mid-infrared continuum between 500 and 1500 cm$^{-1}$.

As $T$ decreases, the infrared response of the Ba122 system evolves in distinct ways depending on the doping level. In the parent compound, the onset of long-range AFM order induces drastic changes of the electronic response. Below the AFM transition temperature $T_N$=135 K, a prominent optical transition associated with the formation of the spin-density-wave (SDW) gap appears in $\sigma_1(\omega)$ between 500 and 1500 cm$^{-1}$. In $\sigma_1(\omega)$ of the OPD Co-Ba122 and P-Ba122 where the long-range AFM order is suppressed, we observe a weak absorption feature at $T \leq 100$ K: a slight upturn in $\sigma_1(\omega)$ at about 500 cm$^{-1}$ followed by a plateau extending to 1500 cm$^{-1}$ as indicated with the arrows in Figs. 1(b) and 1(c). On the other hand, such absorption structure is not evident in $\sigma_1(\omega)$ of OD Co-Ba122 shown in Fig. 1(d).

Before continuing our inquiry into the electrodynamics of Ba122 phases, we stress that the contribution of the mobile carrier to $\sigma_1(\omega)$ in the iron pnictides is not confined to a narrow Drude-like component but extends further in frequency in the form of flat incoherent background [13-20]. This behavior of $\sigma_1(\omega)$ is generic to various classes of correlated electron systems [21]. Our conjecture for the common physical origin of the Drude-like feature and mid-infrared background is supported by the spectral weight (SW) analysis of $\sigma_1(\omega)$ for Co-Ba122 series [13]. The SW contained in $\sigma_1(\omega)$ measures the degree of the renormalization of electronic bandwidth by interactions [21-23]. Only by assigning most of the SW at $\omega$<1500 cm$^{-1}$ to the response of interacting mobile electrons does one find consistency with band structure renormalization revealed by other probes [13]. Since the response below 1500 cm$^{-1}$ is dominated by intra-band processes it is instructive to analyze these data in terms of the frequency-dependent scattering rate $1/\tau(\omega)$. One can obtain $1/\tau(\omega)$ using the extended Drude model (EDM) $1/\tau(\omega) = \omega_p^2/4\pi \, \text{Re}(1/\tilde{\sigma}(\omega))$, where $\omega_p$ is the plasma frequency and $\tilde{\sigma}(\omega)$ is the complex optical conductivity [24]. For consistency of the analysis, we determined the value of $\omega_p$ of each compound from the integration of $\sigma_1(\omega)$ at 295 K up to 1500 cm$^{-1}$. The EDM analysis also requires a correction for higher frequency interband absorption. For this task, we modeled interband transitions using a set of Lorentz oscillators (inset of Fig. 1(e)) and subtracted their contribution from experimental $\tilde{\sigma}(\omega)$. Some uncertainty with the choice of $\omega_p$ or with the particular choice for high-energy corrections will not modify the key outcomes of the EDM analysis. We emphasize that we use the same protocol for the EDM analysis for all the compounds, yet we find markedly different behavior of $1/\tau(\omega)$ at the extremes of the doping level and systematic evolution across the entire doping range.

The scattering rate spectra for Ba122, OPD Co-Ba122, and P-Ba122 (Figs. 1(e)-1(g)) display a number of common trends. As $T$ decreases, $1/\tau(\omega)$ are suppressed below about 700 cm$^{-1}$. For OPD Co-Ba122 and P-Ba122 where the evolution of $1/\tau(\omega)$ can be followed down to $T<10$ K we observe a threshold in $1/\tau(\omega)$ in the region of 600-700 cm$^{-1}$ as indicated by gray bars in Figs. 1(e)-1(g). It is this characteristic frequency dependence that is associated with the pseudogap in the cuprates [2, 24]. The precise form of the absorption threshold in the cuprates is also influenced by coupling to resonant excitations [25]. As we will discuss below, both strong-coupling effects and the pseudogap govern the intra-gap charge dynamics in the Ba122 pnictides. It should be noted that the threshold structure in $1/\tau(\omega)$ of OPD compounds appears at the same frequency region where $\sigma_1(\omega)$ of the parent phase displays the SDW gap. The behavior of the OD crystal is markedly different (Fig. 1(h)). These latter data are dominated by nearly parallel vertical offset of the entire $1/\tau(\omega)$ spectra with the variation in $T$. Neither the pseudogap nor strong-coupling features are needed to account for the gross features of $1/\tau(\omega)$ of the OD crystal.

The structure of $1/\tau(\omega)$ is known to unveil the signatures of interactions in the electromagnetic response of correlated electron systems [21]. Specifically, $1/\tau(\omega)$ data for the cuprates revealed infrared signatures of the superconducting gap at $T<T_c$, of the pseudogap [a partial gap in the density of states (DOS)] in the normal state, and also features originating from electron-boson coupling [24, 25]. The pseudogap with the magnitude $\Delta_{PG}$ leads to a suppression of $1/\tau(\omega)$ at low frequencies followed by a crossover to the regime of strong scattering at $\omega>\Delta_{PG}$ [2, 24]. As a result, a distinct threshold appears in $1/\tau(\omega)$; the frequency range of this threshold quantifies the energy scale associated with the pseudogap [24, 25]. Strong coupling of electrons to resonant bosonic excitations also affects the spectral form of $1/\tau(\omega)$. For example, a sharp mode in the bosonic spectral function $\alpha^2 F(\omega)$ results in a rapid increase in the scattering rate at frequencies above this mode [24, 25]. Sharapov and Carbotte developed a formalism accounting for the concerted effect on $1/\tau(\omega)$ of the pseudogap and of $\alpha^2 F(\omega)$ [26],

$$\frac{1}{\tau(\omega,T)} = \frac{\pi}{\omega}\int_0^\infty d\Omega \alpha^2 F(\omega) \int_{-\infty}^\infty d\varepsilon [N(\varepsilon-\Omega)+N(-\varepsilon+\Omega)]$$

$$\times [n_B(\Omega)+n_F(\Omega-\varepsilon)][n_F(\varepsilon-\omega)-n_F(\varepsilon+\omega)], \qquad (1)$$

where $N(\varepsilon)$ is the normalized DOS, $n_B(\varepsilon)=1/(e^{\varepsilon/kT}-1)$ and $n_F(\varepsilon)=1/(e^{\varepsilon/kT}+1)$ are boson and fermion occupation numbers, respectively.

Most commonly, the strong-coupling analysis is applied to systems where one single band crosses the Fermi surface. In general, multi-band systems can pose caveats for the interpretation of $1/\tau(\omega)$. At least Ba122 family of the pnictides appears to present a special case of a multi-band system where this single-band analysis is applicable and grasps important physics [27]. A meticulous four-band calculation of the optical response for K-doped Ba122 compound points to an overwhelming contribution associated with one single band [27]. The authors also show that the experimental manifestations of the strong-coupling effects in $1/\tau(\omega)$ are remarkably similar to what is usually seen in single-band materials. The underlying reason for predominance of a single band in the infrared data may stem from differences in the effective masses associated with various bands with the "lightest" quasiparticles always dominating the infrared response [28].

To attain detailed information on mechanisms underlying charge dynamics of the iron pnictides, we modeled the experimental $1/\tau(\omega)$ using Eq. (1). For $\alpha^2 F(\omega)$, we employ a Gaussian peak at $100 - 200$ cm$^{-1}$: the energy scale revealed by the neutron studies of magnetic excitations in Ba122 family [29-31]. In order to describe the impact of the pseudogap on the DOS we utilized a quadratic gap function [32],

$$N(\varepsilon) = \left[N(0)+\left((1-N(0))\frac{\varepsilon^2}{(\Delta_{PG}/2)^2}\right)\right]\theta(\frac{\Delta_{PG}}{2}-|\varepsilon|)+\theta(|\varepsilon|-\frac{\Delta_{PG}}{2}), \qquad (2)$$

where $\theta(\varepsilon)$ is the Heaviside function. The calculated spectra of $1/\tau(\omega)$ are shown in Figs. 1(i)-1(l). The right panels of Fig.1 display the corresponding forms of $\alpha^2 F(\omega)$ and DOS employed in the analysis. The principle consequence of the coupling to a sharp mode at $100 - 200$ cm$^{-1}$ in $\alpha^2 F(\omega)$ is a rapid onset of $1/\tau(\omega)$ at low frequencies [33]. The opening of pseudogap results in the suppression of $1/\tau(\omega)$ over much broader frequency range extending to $\sim 700$ cm$^{-1}$. To illustrate the relative roles of the bosonic coupling and of the pseudogap we also calculate the $1/\tau(\omega)$ for P-Ba122 at 32 K by setting $\Delta_{PG}=0$ (Fig. 1(g)). A cursory inspection of the latter spectrum shows a rapid increase of $1/\tau(\omega)$ due to the resonance mode that is at variance with the gradual character of the experimental data. Furthermore, our modeling confirms that the impact of bosonic coupling on the charge dynamics is confined to low frequencies thus attesting to the notion that a pseudogap in the DOS is needed to account for a gradual threshold structure at 700 cm$^{-1}$ in our data.

A notable outcome of the analysis reported in Fig.1 is the correlation between the evolutions of the resonant mode and of the pseudogap. The sharp mode in $\alpha^2F(\omega)$ of parent and OPD compounds becomes progressively weaker with increasing $T$, concomitant with the reduction of the pseudogap depth. This correlation also holds for the OD Co-Ba122: the mode in $\alpha^2F(\omega)$ is very weak and there is no need to invoke a pseudogap to explain $T$- or $\omega$-dependences of $1/\tau(\omega)$. All these results suggest that the electronic pseudogap and the resonance mode may be of common microscopic origin ultimately related to magnetism in the iron pnictides as we will discuss later. The $1/\tau(\omega)$ analysis therefore demonstrates that infrared properties of both parent and OPD compounds are dominated by the pseudogap with the magnitude $\Delta_{PG}\approx700$ cm$^{-1}$ that develops at $T$ exceeding $T_N$ of Ba122; the pseudogap vanishes on the OD side of the phase diagram.

Having identified the normal-state pseudogap, we now assess its relationship to the superconducting gap $2\Delta_{SC}$ of P-Ba122 ($T_c$=30 K) and OPD Co-Ba122 ($T_c$=22 K). An infrared signature of the energy gap in dirty superconductors is the onset of absorption at $\omega=2\Delta_{SC}$ followed by a rapid increase of $\sigma_1(\omega)$. This behavior is indeed observed in OPD Co-Ba122 (Fig. 2(b)) where we estimate $2\Delta_{SC}\approx50$ cm$^{-1}$ [34]. In clean superconductors the onset of absorption is expected to start at $4\Delta_{SC}$ [35]. In the intermediate regime, a shallow onset of the conductivity can be recognized at $2\Delta_{SC}$, whereas the maximum of $\sigma_1(\omega)$ occurs near or above $4\Delta_{SC}$ [35]. This latter behavior is consistent with the response of P-Ba122 where we find a gradual increase of $\sigma_1(\omega)$ between 80 and 200 cm$^{-1}$ (Fig. 2(a)). We note that impurity scattering in P-Ba122 is likely to be weaker [12, 36] compared to Co-Ba122 since the P dopants leave the conducting Fe planes intact. The above interpretation of the electrodynamics of P-Ba122 is in accord with our $1/\tau(\omega)$ data: the value of $1/\tau(\omega\rightarrow0)$ at $T\sim T_c$ of P-Ba122 is smaller than that of OPD Co-Ba122 by a factor of 3. Further support for the clean-limit scenario comes from quasilinear $T$ dependence of the change in the magnetic penetration depth $\Delta\lambda$ of P-Ba122 [36]. On the other hand, $\Delta\lambda$ of Co-Ba122 compounds shows a power-law behavior [37]; $\Delta\lambda\sim T^n$ with the exponent $n$ larger than 2. This behavior is consistent with the notion of the substantial pairbreaking impurity scattering within the $s_\pm$-wave scenario [37, 38]. The value of $2\Delta_{SC} \approx 80$ cm$^{-1}$ of P-Ba122 inferred from Fig. 2(a) is in reasonable agreement with that from recent photoemission spectroscopy experiments [39]. Finally, we remark that Ba$_{1-x}$K$_x$Fe$_2$As$_2$ crystals are also likely to be less disordered than Co-Ba122 and display a gradual increase of $\sigma_1(\omega)$ above $2\Delta_{SC}$ [20, 27] similar to P-Ba122.

Our analysis of the electromagnetic response above and below $T_c$ allows us to comment on the possible origins of the pseudogap state in the iron pnictides. Experiments reported in Fig. 1 reveal that the magnitude of the normal-state pseudogap significantly exceeds that of the superconducting gaps. This finding is at odds with a "precursor to superconductivity" interpretation of the pseudogap in the iron pnictides. A similar conclusion holds for the cuprates where the two gaps are also well separated in energy in the underdoped or OPD compounds [1] and where the pseudogap was attributed to a non-superconducting broken-symmetry state [40].

The appearance of the pseudogap in the electronic response of Ba122 system correlates with the evolution of the AFM fluctuations across the phase diagram. Neutron scattering experiments on Ba122 determined that the AFM fluctuations related to the SDW instability persist above $T_N$ [41]. NMR studies found that the AFM fluctuations in the parent compound remain robust in P-Ba122 and OPD Co-Ba122 but become drastically suppressed in OD Co-Ba122 [3, 42]. Furthermore, the energy scale of the infrared pseudogap in OPD compounds is in accord with that of the SDW gap of the parent compound. The strong-coupling analysis of $1/\tau(\omega)$ further reveals the intimate correlation between the development of the resonance mode of $\alpha^2F(\omega)$ and the pseudogap. Thus the totality of data suggests that the AFM fluctuations related to the SDW instability of the parent compounds might be a plausible cause for the pseudogap of the iron pnictides. Recent experiments showed that the AFM and orbital fluctuations can coexist [43, 44]. Plausibly, both these effects may be important for the pseudogap formation.

Both the spectral manifestations of the pseudogap in the pnictides and the evolution of these features across the phase diagram are strikingly similar to those of the hole-doped cuprates. The manifestation of the pseudogap in $\sigma_1(\omega)$ of the hole-doped cuprates is subtle: a shallow dip in otherwise nearly featureless spectra $\sigma_1(\omega)$ [2, 24]. Very similar behavior is also observed in the conductivity data of the iron pnictides. The infrared signature of the pseudogap is much more obvious in the electron-doped cuprates [45]. We stress that the prominence of AFM fluctuations is one of the very few unifying aspects of these two classes of high-$T_c$ superconductors: the cuprates and the pnictides.

A salient feature of the pseudogap state in the cuprates is nematic correlations leading to the anisotropic response of their building blocks: $CuO_2$ planes [46, 47]. Recent experiments on Co-Ba122 have also identified in-plane transport anisotropy attributed to nematic effects [48, 49] in the parameter space where our infrared data detect the pseudogap. Moreover, thermodynamic and ultrasound spectroscopic studies indicate that a nematic state exists well above $T_N$ and persists to the nonmagnetic superconducting regime [50, 51]. Further connections between the pseudogap characteristics of the two classes of materials need to be explored in more details. A tentative conclusion based upon the data presented in Fig. 1, as well as the literature, is that the AFM state of parent materials has a direct bearing to the pseudogap in both classes of high-$T_c$ superconductors.

We thank S. L. Bud'ko for his assistance with the sample growth and characterization. S.J.M., A.A.S., and D.N.B. acknowledge support from the National Science Foundation (NSF 1005493) and AFOSR. S.J.M. acknowledges support from Basic Science Research Program through the National Research Foundation of Korea funded by the Ministry of Education, Science, and Technology (2012R1A1A1013274). The work at Ames Laboratory was supported by the U.S. Department of Energy, Office of Basic Energy Science, Division of Materials Sciences and Engineering under Contract No. DE-AC02-07CH11358. Work at ORNL was supported by the U.S. Department of Energy, Basic Energy Sciences, Materials Sciences and Engineering Division.

[*]dbasov@physics.ucsd.edu

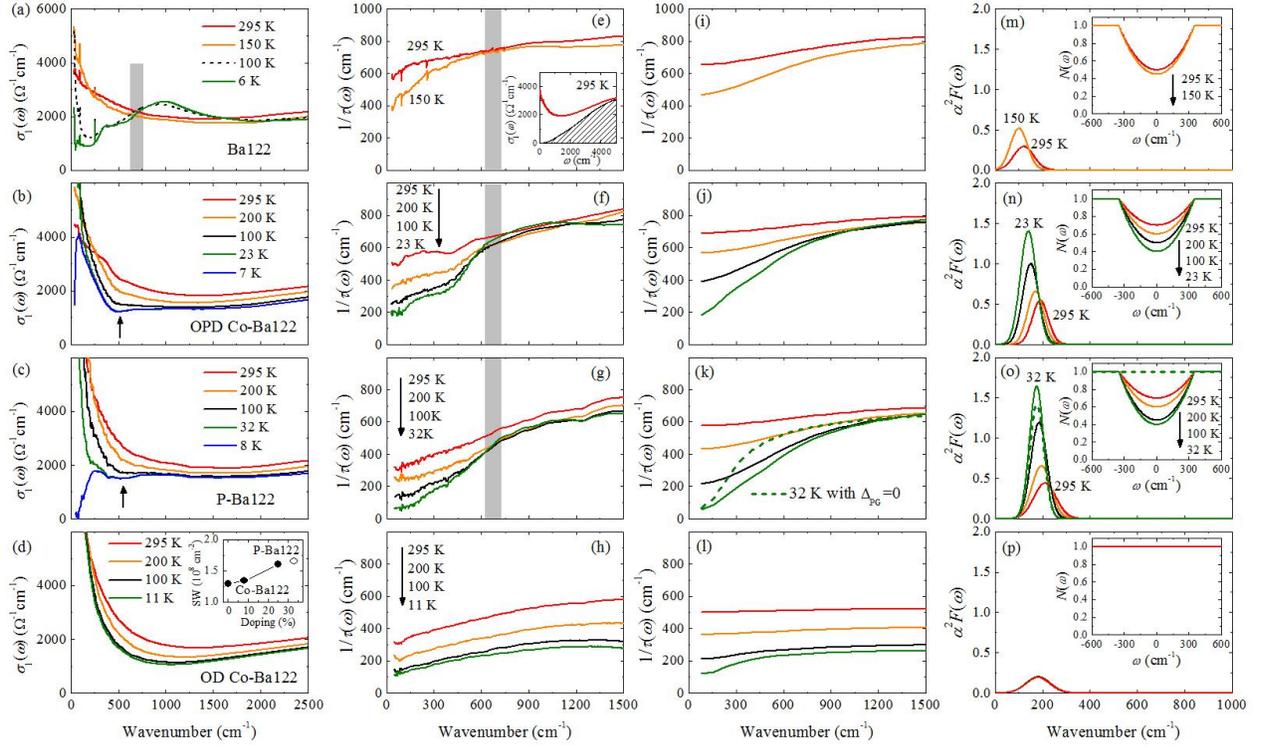

FIG.1 (color online). (a)-(d) $T$-dependent $\sigma_1(\omega)$ of BaFe$_2$As$_2$, OPD Ba(Fe$_{0.92}$Co$_{0.08}$)$_2$As$_2$, BaFe$_2$(As$_{0.67}$P$_{0.33}$)$_2$, and OD Ba(Fe$_{0.75}$Co$_{0.25}$)$_2$As$_2$ from top to bottom. Inset of (d): spectral weight obtained by integrating $\sigma_1(\omega)$ at 295 K up to 1500 cm$^{-1}$. The gray bar indicates the SDW gap in BaFe$_2$As$_2$. The arrows in (b) and (c) denote the frequency above which $\sigma_1(\omega)$ displays slight increase due to the pseudogap. (e)-(h) Scattering rate $1/\tau(\omega)$. The gray bars represent the frequency range below which $1/\tau(\omega)$ exhibits suppression in the magnitude. Note that this frequency region correlates with the SDW gap in $\sigma_1(\omega)$ of BaFe$_2$As$_2$. Inset of (e): the hatched area denotes the contribution from interband transitions. (i)-(l) Calculated $1/\tau(\omega)$ spectra. The dashed line in (k) shows the calculation result for $1/\tau(\omega)$ at 32 K assuming vanishing pseudogap ($\Delta_{PG}$=0). (m)-(p) $\alpha^2F(\omega)$ spectra and DOS (insets) obtained from the $1/\tau(\omega)$ analysis. The dashed lines in (o) and its inset represent $\alpha^2F(\omega)$ and DOS for $1/\tau(\omega)$ result with $\Delta_{PG}$=0, respectively.

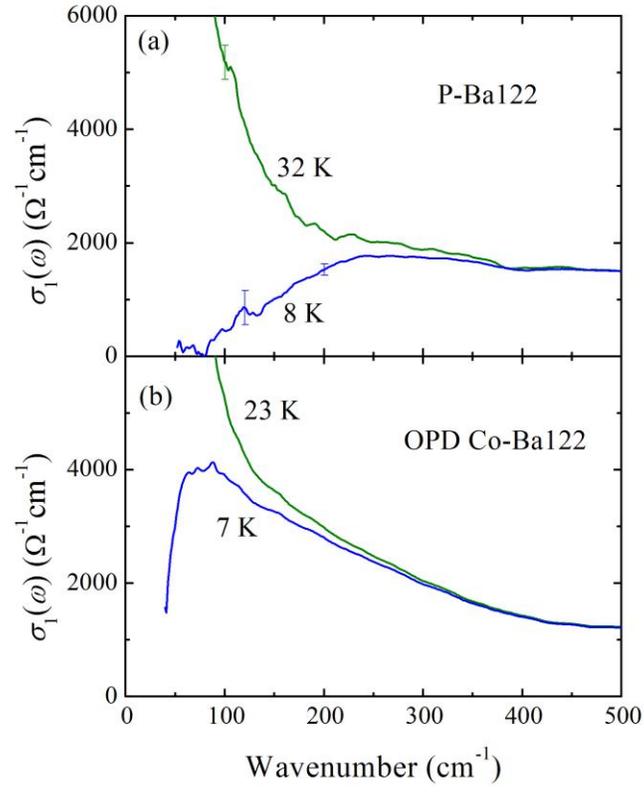

FIG. 2 (color online). $\sigma_1(\omega)$ in the far-infrared region. (a) BaFe$_2$(As$_{0.67}$P$_{0.33}$)$_2$. Vertical bars on top of $\sigma_1(\omega)$ represent error bar. (b) OPD Ba(Fe$_{0.92}$Co$_{0.08}$)$_2$As$_2$.

## Supplemental Material

**Experimental**

High-quality single crystals of BaFe$_2$(As$_{0.67}$P$_{0.33}$)$_2$ (P-Ba122) and Ba(Fe$_{1-x}$Co$_x$)$_2$As$_2$ (Co-Ba122) were grown using self-flux method. The composition of the crystals was determined by an energy dispersive x-ray spectroscopy analysis. Details of growth procedures and characterization results are described in Refs. S1-S4. Parent compound Ba122 ($x$=0) exhibits structural and magnetic transition at $T_N$=135 K [S1]. Optimally doped (OPD) Co-Ba122 ($x$=0.08) and P-Ba122 show superconducting transition at $T_c$=22 and 30 K, respectively [S1, S2]. Overdoped (OD) Co-Ba122 ($x$=0.25) is nonsuperconducting [S4]. The $ab$-plane dimensions of the crystals used for the infrared experiments are following: Ba122 (2×2 mm$^2$), OPD Co-Ba122 (4×4 mm$^2$), P-Ba122 (1×1 mm$^2$), and OD Co-Ba122 (4×4 mm$^2$).

The $ab$-plane reflectance $R(\omega)$ of these crystals were measured at various temperatures using in-situ overcoating technique [S5]. The complex optical conductivity $\tilde{\sigma}(\omega) = \sigma_1(\omega) + i\sigma_2(\omega)$ was determined from the Kramers-Kronig analysis of $R(\omega)$ [S6].

**Reflectance Spectra**

Figure S1 displays temperature-dependent $R(\omega)$ of the Ba122 family. Low-frequency $R(\omega)$ of all the compounds shows a metallic behavior as attested by the large magnitude and its increase with decreasing temperature.

In the parent compound Ba122, $R(\omega)$ experiences drastic changes associated with the spin-density-wave gap formation in the antiferromagnetic state. In OPD Co-Ba122 and P-Ba122 compounds where the long-range antiferromagnetic order is suppressed, we observe a weak downward kink near 500 cm$^{-1}$, which becomes stronger at low temperatures. It is this feature that is related to the pseudogap identified in the real part of optical conductivity $\sigma_1(\omega)$ and frequency-dependent scattering rate $1/\tau(\omega)$ shown in the main paper. By contrast, as temperature decreases, $R(\omega)$ of OD Co-Ba122 shows continuous upward vertical shift without exhibiting such kink feature.

The onset of superconductivity is clearly observed in raw $R(\omega)$ of P-Ba122 and OPD Co-Ba122. Below $T_c$, $R(\omega)$ of P-Ba122 (OPD Co-Ba122) exhibits sudden upturn at about 200 cm$^{-1}$ (100 cm$^-$

$^1$) and approaches unity at about 80 cm$^{-1}$ (50 cm$^{-1}$), which is a strong indication of the formation of superconducting energy gap [S7].

**Superfluid Density of P-Ba122 and OPD Co-Ba122**

The superfluid density $\rho_s$ can be estimated from the imaginary part of optical conductivity $\sigma_2(\omega)$ in the zero-frequency limit: $\rho_s=4\pi\omega\sigma_2(\omega\rightarrow 0)$. As shown in Fig. S2, this analysis yields $\rho_s\approx 8.1\times 10^7$cm$^{-2}$ for P-Ba122, which agrees well with the result of microwave measurements[S8]. The magnitude of $\rho_s$ of OPD Co-Ba122 is found to be about $2.2\times 10^7$cm$^{-2}$. The large difference in the magnitude of $\rho_s$ between these two compounds can be due to the variation in the concentration of pairbreaking impurities. Indeed, thermodynamic studies showed that P-Ba122 is close to the clean limit [S8], whereas Co-Ba122 experiences substantial paribreaking impurity scattering [S9, S10].

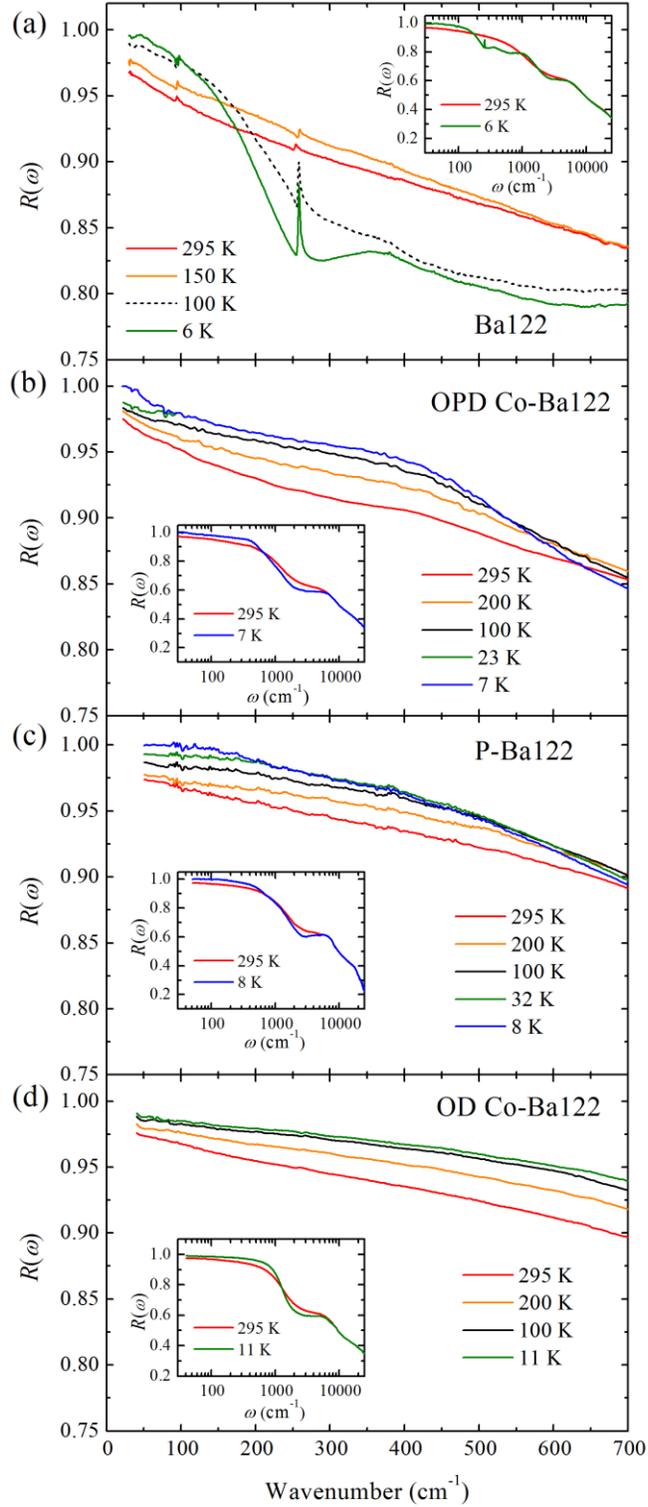

FIG. S1. Temperature-dependent reflectance spectra $R(\omega)$. (a) Ba122, (b) OPD Co-Ba122, (c) P-Ba122, and (d) OD Co-Ba122. Insets show $R(\omega)$ in broad frequency range up to 25000 cm$^{-1}$.

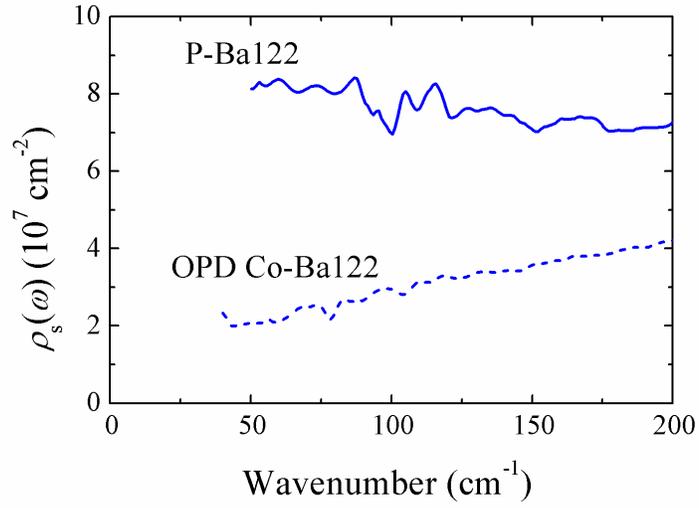

FIG. S2. Superfluid density $\rho_s(\omega)$ of P-Ba122 and OPD Co-Ba122 obtained from imaginary part of optical conductivity $\sigma_2(\omega)$: $\rho_s = 4\pi\omega\sigma_2(\omega)$.